\documentclass[aps,prl,twocolumn,groupedaddress,letterpaper,showpacs]{revtex4}
\usepackage{graphicx}

\begin{document}


\title{Intensity Interferometry with Anyons}


\author{Thomas D. Gutierrez}
\email{tgutierr@lifshitz.ucdavis.edu}
\homepage{http://nuclear.ucdavis.edu/~tgutierr}
\affiliation{Department of Physics, University of California, Davis}


\date{\today}

\begin{abstract}

A pairwise correlation function in relative momentum space is  
discussed as a tool to characterize the properties of an incoherent source of non-interacting Abelian anyons. 
This is analogous to the Hanbury--Brown Twiss effect for particles with fractional statistics in two dimensions.
In particular, using a flux tube model for anyons, the effects of source shape and quantum statistics on a 
two-particle correlation function are examined.  
Such a tool may prove useful in the context of quantum computing and other experimental
applications where studying anyon sources are of interest.
\end{abstract}


\maketitle

Anyons are particles that exhibit fractional quantum statistics in two dimensions.
Existing somewhere between 
bosons and fermions, anyons have been applied theoretically 
to a variety of problems such as the fractional 
quantum Hall effect (FQHE) \cite{fqhhal84k,fqharo84}, high temperature
superconductivity \cite{alau88},
supersymmetry
\cite{suply97,surau97}, and fault-tolerant quantum computing \cite{qckit03}.  
Experimentally, the quasi-particle excitations
seen in the FQHE have been shown to possess anyon properties \cite{aclar88,asim89}.

Fractional spin in two dimensions directly addresses the topological interpretation
of quantum statistics.
For identical particles in three or more dimensions, the appropriate homotopy group is the permutation group \cite{homwu84}  
and there are only two homotopy classes. These correspond directly
to fermions and bosons.  

However, in two dimensions, the corresponding homotopy group is the braid group \cite{homwu84}
and an infinite number of homotopy classes are possible.  
As particles are exchanged in
relative position, each successive winding of the particles around each other cannot be smoothly deformed 
into a finite number of configurations as is the
case in higher ($>$2) dimensions.  
Each winding corresponds to a separate homotopy class which in turn can be categorized as a different
classification of quantum statistics.  
The term {\em anyon} was coined by Wilczek \cite{awil4882,awil4982} to describe objects having such topological properties,
a physical example of which is a charged particle bound to a magnetic flux tube in two dimensions.  

This relationship between topology and statistics was 
glimpsed by Aharonov and Bohm \cite{ab} and later realized,
formalized, and expanded by others 
\cite{alein77,agold80,agold81,awil4882,awil4982,afor92,akhare1}.  
Because the homotopy group is the braid group, this makes many
multi-particle ($>$2) anyon problems nearly intractable.
An essential analytical difficulty arises because, unlike the case with the permutation group, 
the multi-particle wave functions for anyons cannot
in general be written in a simple way in terms of the single-particle wave functions.  
Multi-particle anyon wave functions are, in effect, permanently entangled.  
However, it is exactly this topological property of anyons that makes
them robust against decoherence and thus desirable candidates for qubits in fault-tolerant quantum computing \cite{qckit03}.  

Because quantum interference effects can be 
sensitive to quantum statistics, it is natural to inquire about the role
of anyons in this context.  As mentioned, the Aharonov-Bohm effect was an early probe into
this fascinating problem.  More recently, first order interference 
effects in a Mach-Zender-style interferometer
for non-Abelian anyons in the presence of Aharonov-Bohm flux sources have been
derived from applications of the braid group \cite{aover01}.  

In this work, intensity interferometry in momentum space, 
a second-order interference effect, is explored for incoherently emitted, non-interacting Abelian anyons.
This is conceptually related to the recent q-Bose gas interferometry approaches for pions 
in heavy ion collisions \cite{qbanch00a,qbanch00b}.  
However, q-bosons and q-fermions are not restricted to two dimensions. 
In that case, the generalization of quantum statistics is achieved
somewhat differently than for anyons \cite{qhald91}. 

Intensity interferometry is a useful tool for characterizing the source geometry of incoherently emitted particles.  
In addition, the interference effect, expressed in terms
of a second-order correlation function, might be regarded as an
entanglement measure.  Such a tool may prove useful in the context of quantum computing and other 
experimental applications where studying anyon sources are of interest.  

Intensity interferometry was
originally developed by Hanbury--Brown and Twiss (HBT)
as an alternative to Michelson interferometry to measure the angular sizes of stars
in radio astronomy \cite{hbt}.
By correlating intensities rather than amplitudes, the measurement is
insensitive to high frequency fluctuations that would normally make Michelson
interferometry prohibitive.  When applied to classical waves, 
the HBT effect is essentially a beat phenomenon.

The field of modern quantum optics was spawned when intensity interferometry was quantum mechanically
applied to photons rather than classical waves.
Now, two- and multi-photon effects are routinely
studied in photonics.  Two-fermion HBT in 2D condensed matter systems
has also been reported \cite{hbthen99}.
The HBT effect was independently
applied to pions and other particles in elementary particle
physics in momentum space
and is often called the Goldhaber-Goldhaber-Lee-Pais (GGLP) effect in that context 
\cite{hbtgold60,hbtboal90,hbtbay98}.

Quantum mechanically, the two-particle correlation function is 
a measure of the degree of independence between a joint measurement of two 
particles ($1$ and $2$ below).  Formally, the 
correlation function can be written:

\begin{equation}
C_2=\frac{{\rm Tr}[\rho a^{\dagger} a^{\dagger} a a]}{{\rm Tr}[\rho a^{\dagger} a]{\rm Tr}[\rho a^{\dagger} a]}\sim\frac{P(1,2)}{P(1)P(2)}
\label{c2}
\end{equation}
where $\rho$ is the density matrix and $a^{\dagger}$ and $a$ are the creation and annihilation 
operators for the quanta associated with the appropriate fields of interest.  
When normalized to the single particle distributions as shown,  $C_2$ is proportional to the relative probability for
a joint two-particle measurement as compared to two single-particle measurements.  If the measurements
are independent, then $C_2$=1.  If the measurements are correlated, $C_2$ deviates from unity.
In principle, $C_2$ can be expressed as a joint measurement as a function of any degree of freedom of interest.
In this treatment, we will focus on momentum space correlations.  

The second-order correlation function in momentum space is a powerful tool to probe several important properties
of a system.  $C_2$ is sensitive to the quantum statistics of the particles as expressed in the commutation relations for
the field operators in Eq.~(\ref{c2}).  Also, as determined by the form of the density matrix, $C_2$ contains information
about the space-time distribution of the particle source in phase space as well as the pairwise 
interaction and the quantum field configuration.  

Non-interacting identical particles can
exhibit strong correlations, no correlations, or even anti-correlations depending
on the specific field configuration.  Possible field configurations for identical bosons include 
thermal states, coherent states, Fock states, and squeezed states \cite{qoscull1}.  For example, an 
incoherent thermal field configuration gives $C_2=2$ for spinless bosons of equal momentum.   
In contrast, a coherent source of bosons, such as laser light well above the lasing threshold, gives a constant $C_2=1$ 
for {\em all} relative momenta.

Geometric information about the source is contained in the shape of the correlation function
as a function of the pair's relative momentum.  
The correlation function typically approaches unity for large momentum differences.  
For incoherent sources, the relationship between source width and correlation width is essentially limited
by the Heisenberg uncertainty principle such that $\delta q\sim 1/R$ where $\delta q$
is the width of $C_2$ in relative momentum space and $1/R$ is some characteristic
width of the incoherent source generating the pairs.

Keeping in mind the above discussion, this paper examines
the behavior of $C_2$ in momentum space for a non-expanding, 
non-relativistic, incoherent source of anyons.  
An approximation
to Eq.~(\ref{c2}) is used to highlight some features of intensity interferometry in momentum
space. The quantum statistics are ``tuned'', and, given some simple source functions, the 
corresponding correlation function is extracted.

In momentum space for an incoherent source, Eq.~(\ref{c2}) reduces to 
the Koonin-Pratt equation \cite{hbtkoo77,hbtpra90,hbtbro97}

\begin{equation}
C_2({\bf q})-1=\int d{\bf r} K({\bf q},{\bf r}) S({\bf r})
\label{kp}
\end{equation}
where the integration kernel is given by

\begin{equation}
K({\bf q},{\bf r})=|\Phi_{{\bf q}}({\bf r})|^2-1.
\label{kern}
\end{equation}

The function $\Phi_{{\bf q}}({\bf r})$ is the two-particle wave function in the center of mass frame of
the pair, where ${\bf q}=({\bf p_1}-{\bf p_2})/2$ is the relative momentum and ${\bf r}={\bf r_1}-{\bf r_2}$ is the 
relative separation in that frame.  The center of mass motion of the pair is not considered in this
treatment.  The source function,
$S({\bf r})$, is the normalized probability distribution of emitting 
a particle pair with relative separation ${\bf r}$.  The integral is over the entire relative
separation space.  In this context, incoherent means that the particles are emitted from the source randomly
and independently.  
Moreover, any potential time dependence of the source function in Eq.~(\ref{kp}) has been integrated out 
as in Ref.~\cite{hbtbro97}.
This integration eliminates the ability to recover emission-time ordering information between the pairs and 
places the focus on 
spatial information only.

Using a flux tube model for anyons,
a dynamical approach is used to obtain two-particle non-relativistic
wave functions for anyons in two dimensions 
\cite{ab,awil4882,awil4982,akhare1}.  
In this model, the quantum statistics are enforced through an interaction, the strength of which 
enforces the effective quantum statistics.
In this picture, each anyon is analogous to a charged particle bound to a tightly bundled magnetic field
in two-dimensions.  The analogy is not strictly one-to-one in this treatment because 
the scalar Coulomb potential between the charged pairs is not considered; indeed, the system need
not be electromagnetic.  Nevertheless, the analogy 
paints a helpful physical picture.

The tightly bundled magnetic field associated with each anyon, $\vec{B}=\nabla\times\vec{A}$, 
points perpendicular to the plane of motion.  We can select a gauge such that
the vector potential, $\vec{A}$, has the form $A_r=0$ and $A_{\phi}=\xi/2\pi r$, where $\xi$ is the
magnetic flux through the plane.  The particles are only permitted 
in regions of space where the magnetic field of the other
particle is zero, so
the particles exert no force on each other.
Quantum mechanically, the interaction term takes the form of a minimal coupling such that the 
momentum operator takes the form $\hat{p}\rightarrow(-i \nabla-e\vec{A})$.  
While under the conditions described above, this ``interaction'' exerts no
force, it does adjust the phase of the wave functions, permitting interference effects.
The relative wave function of the pair
can then be used in Eq.~(\ref{kp}), in combination with a normalized source distribution, to
obtain $C_2$.  

In the center-of-mass frame, using the Hamiltonian 

\begin{equation}
\hat{H}=\frac{\hat{p}_r^2}{m}+\frac{(\hat{p}_\phi-\alpha)^2}{m r^2},
\label{anyHam}
\end{equation}
the wave function is obtained 
for the relative motion of two free anyons in two-dimensional polar coordinates.
In Eq.~(\ref{anyHam}), $m$ is the mass of one particle ($m_{1}=m_{2}=m$) and $\alpha$ represents the anyon parameter
with $0\le\alpha\le1$, where $\alpha=0 (1)$ corresponds to bosons (fermions).  In this standard treatment,
all of the details of the vector potential, magnetic flux, and gauge choice are included in 
the tunable parameter $\alpha$.  In Eq.~(\ref{anyHam}), the
reduced mass of the identical pair, $\mu=m_{1}m_{2}/(m_1+m_2)=m/2$, has already been substituted.
In the two-dimensional relative coordinate system of the identical pairs, the x-axis has been chosen along the direction
of $\vec{q}$.

Applying Eq.~(\ref{anyHam}) to 
the time-independent Schr\"odinger equation, $\hat{H}\psi=E\psi$, gives 

\begin{equation}
[(\frac{\partial^2}{\partial r^2}+\frac{1}{r}\frac{\partial}{\partial r})-\frac{1}{ r^2}(i
\frac{\partial}{\partial\phi}+\alpha)^2+q^2]\psi=0
\label{anySE}
\end{equation}
where $q^2=m E$ is the square of the relative momentum of the pair 
(i.e. the energy is that of a free relative particle) and $\phi$ is the angle between
${\bf q}$ and ${\bf r}$.  
Eq.~(\ref{anySE}) admits solutions of the form

\begin{equation}
\psi(r,\phi)=N J_{|l-\alpha|}(qr)e^{il\phi}
\label{sol1}
\end{equation}
where $N$ is a normalization constant.
We want solutions that correspond to the appropriate 
symmetric (antisymmetric)
boson (fermion) free particle wave function in relative coordinates
in the limit of $\alpha=0 (1)$.  
Specifically,

\begin{equation}
\Phi_{\alpha=0,1}(r,\phi)=\frac{1}{\sqrt{2}}(e^{i{\bf q}\cdot{\bf r}}\pm e^{-i{\bf q}\cdot{\bf r}})
\label{bosfer}
\end{equation}
where the $\alpha=0$ case corresponds to the symmetric solution (+) while the $\alpha=1$ case is the
antisymmetric solution (-).

Keeping in mind the identity
\begin{equation}
e^{i{\bf q}\cdot{\bf r}}=e^{qr\cos{\phi}}=\sum_{n=-\infty}^{+\infty}i^n J_n(qr) e^{in\phi}
\label{bessid}
\end{equation}
one can construct partial wave superpositions of Eq.~(\ref{sol1}) that give Eq.~(\ref{bosfer})
for the limiting cases of $\alpha=0,1$.  Namely,

\begin{equation}
\Phi_{q,\alpha}(r,\phi)=\sqrt{2}\sum_{l=-\infty (\rm{even})}^{+\infty}
(i)^{|l-\alpha|}J_{|l-\alpha|}(qr)e^{il\phi} 
\label{anywf}
\end{equation}
where $l$ is
the orbital angular momentum quantum number.  The sum is over partial waves of Bessel functions of the first kind of 
order $|l-\alpha|$ where $l$ is necessarily an integer, but $\alpha$ is not.  
The subscript $q$ has been added to the wave function at this stage 
to remind the reader that
we will be integrating over $r$ and $\phi$ using Eqs.~(\ref{kp}) and (\ref{kern}) to obtain a function of $q$.

Treating the initial pair wave function in Eq.~(\ref{anySE}) as that of spinless bosons (the ``boson basis'')
simplifies the problem of obtaining the appropriate 
exchange symmetry for the final two-particle wave function shown in Eq.~(\ref{anywf}).  In constructing
Eq.~(\ref{anywf}), the partial wave expansion using Eq.~(\ref{sol1})
is symmetrized under the exchange $\phi\rightarrow(\phi-\pi)$. 
Tuning the anyon parameter will result in a wave function that possesses the
correct exchange symmetry for identical indistinguishable pairs. For the fermion and boson cases,
Eq.~(\ref{anywf}), with the help of Eq.~(\ref{bessid}), reduces to Eq.~(\ref{bosfer}).  But for
$0<\alpha<1$, Eq.~(\ref{anywf}) is the non-trivial two-anyon free particle wave function.

Equation (\ref{kp}) is used to obtain an expression of $C_2$ for anyons.  While intensity interferometry
can be used to image three dimensional sources, for simplicity we will only consider angle-averaged sources
that are a function of $r$ rather than ${\bf r}$.  
The integration kernel in Eq.~(\ref{kern}) is angle-averaged in two dimensions
and then denoted $K_{0_{\alpha}}(q,r)$ such that

\begin{equation}
K_{0_{\alpha}}(q,r)
=[\frac{1}{2\pi}\int_{0}^{2 \pi} d\phi|\Phi_{q,\alpha}(r,\phi)|^2]-1.
\label{aakern}
\end{equation}

Inserting the expression for the relative wave function from Eq.~(\ref{anywf}) into Eq.~(\ref{aakern}) and performing
the angle-averaging in two dimensions gives

\begin{equation}
K_{0_{\alpha}}(q,r)=[2 \sum_{l=-\infty (\rm{even})}^{l=+\infty} |J_{|l-\alpha|}(qr)|^2]-1.
\label{akern}
\end{equation}

In the limiting case when $\alpha=(0,1)$, Eq.~(\ref{akern}) simplifies to, with the help of
Bessel function properties,

\begin{equation}
K_{0_{\alpha=0,1}}(q,r)=(-1)^{\alpha} J_{0}(2qr)
\end{equation}
which is, by construction, the same angle-averaged kernels one gets by using Eq.~(\ref{bosfer}) directly.
For the more general case when $0<\alpha<1$, the sum in Eq.~(\ref{akern}) is evaluated numerically to an
appropriately converging order.


Two sample source functions are used to illustrate the effects of source shape and size on $C_2$:

\begin{equation}
S_g(r)=\frac{e^{-r^2/4r_0^2}}{4\pi r_0^2}
\label{gsource}
\end{equation}

\begin{equation}
S_s(r)=\frac{\Theta(r_0-r)}{\pi r_0^2}.
\label{esource}
\end{equation}

They represent incoherent pair emission distributions normalized in two dimensions,

\begin{equation}
2 \pi \int_{0}^{\infty} r dr S(r)=1,
\end{equation}
and where the effective source width is given by $r_0$.  The forms of the functions were chosen 
as examples to represent typical localized incoherent sources.  

Integrating Eq.~(\ref{kp}) over the azimuthal angle
in two dimensions using Eq.~(\ref{akern}), and choosing either Eqs.~(\ref{gsource}) or~(\ref{esource})
as the source function,
gives an angle-averaged expression for $C_2$ parameterized by $\alpha$:

\begin{equation}
C_{2_{\alpha}}(q)-1=2\pi\int_{0}^{\infty} r dr K_{0_{\alpha}}(q,r) S(r).
\end{equation}

\begin{figure}[t]
\resizebox{.45\textwidth}{!}{\includegraphics{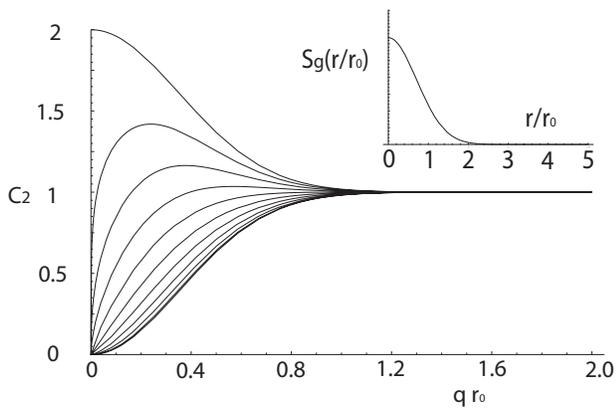}}
\caption{\label{anyoncors} 
$C_{2_\alpha}$ versus $q r_0$ from a normalized Gaussian source given by Eq.~(\ref{gsource}) 
in two dimensions.  The curves represent different values of the anyon parameter $\alpha$.  
The top curve is the boson case ($\alpha=0$) and the bottom curve is the fermion case ($\alpha=1$).  From top to bottom,
$\alpha=$0, 0.1, 0.2, 0.3, 0.4, 0.5, 0.6, 0.7, 0.8, 0.9, and 1.  The source function, plotted in 
arbitrary units, is shown inset versus $r/r_0$.
}
\end{figure}

Figure~\ref{anyoncors} shows $C_{2_{\alpha}}$ for an incoherent Gaussian source, Eq.~(\ref{gsource}).  
Various values of $\alpha$ are displayed.
The figures are shown plotted versus the quantity $q r_0$, which accounts for the shifting
momentum scale upon changing the source size.  For example, as the width of the source decreases, 
the correlation function widens and vice versa.  

For the boson and fermion cases, the standard HBT results for incoherent sources are recovered.
The top curve in Fig.~\ref{anyoncors} represents the spin zero case and approaches a value of
$C_{2_{\alpha=0}}=2$ as $q\rightarrow0$.  Here, it is more likely to measure a pair of non-interacting 
identical bosons in a state of zero relative
momentum than to independently measure each particle with their respective momentum.  That is, in the 
joint measurement, the bosons are not independent.  This effect is a function of $q$ and
is the result of the Bose-Einstein statistics and the symmetric nature of the wave function. 
Similarly, the lowest curve represents non-interacting indistinguishable fermions and displays an anti-correlation 
in the joint measurement at low relative momentum, where $C_{2_{\alpha=1}}=0$ as $q\rightarrow0$.
Similar to the boson case, this is also only due to the quantum statistics and indicates again that a measurement
of one particle affects the other quantum mechanically.  
As the relative momentum increases, the value of $C_{2_{\alpha}}$
approaches unity, indicating that the joint measurement becomes less correlated.

Starting from the top boson curve and moving downward are the correlation curves 
for various values of the anyon parameter.  
Like fermions, the anyons display an exclusion principle and always anti-correlate in the limit 
as $q\rightarrow0$.
However, other than the exclusion at $q=0$, there is a natural trend of decreasing correlation
as one interpolates between bosons and fermions for values of $q>0$.  The correlation function also
approaches unity for large values of $q$. 

\begin{figure}[]
\resizebox{.45\textwidth}{!}{\includegraphics{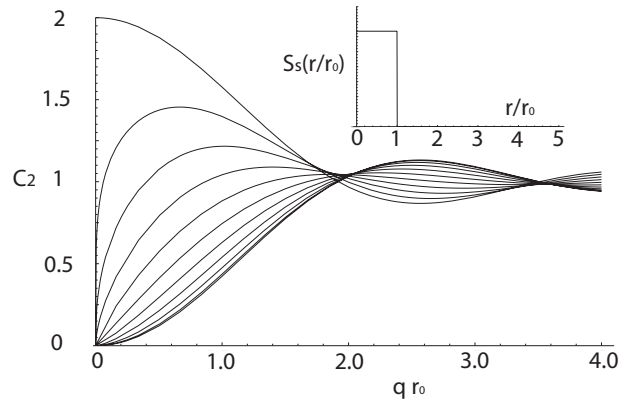}}
\caption{\label{anyoncorsExp} $C_{2_\alpha}$ from a normalized step function source given by Eq.~(\ref{esource}).  
The anyon parameterization is the same as in Fig.~\ref{anyoncors}. 
The source function, plotted in arbitrary units, is inset plotted versus $r/r_0$.}
\end{figure}

An example of the sensitivity of $C_{2_\alpha}$ to the source shape is shown in Fig.~\ref{anyoncorsExp}.  The box profile
with a sharp edge introduces a ringing structure into the correlation function.

Experimentally, the measured correlation function provides information about the size and shape
of the emission source.  Measured
correlation functions can be inverted to determine actual source 
distributions using methods like those discussed in Ref.~\cite{hbtbro97}.
Also, information about anyon pairwise interactions can be extracted by comparing measured correlation functions
against predictions made using hypothesized interactions in the Hamiltonian.
Moreover, the degree of chaoticity of the source, testing the independent emission assumption, can be extracted by 
looking at the behavior
of the correlation at low relative momentum.  Finally, the correlation function itself provides information about the 
degree of entanglement between 
pairs versus their relative momentum, a potentially useful tool in the field of quantum computing.

In summary, intensity interferometry in momentum space 
provides a tool that can be used to experimentally study the properties of anyons
and their sources in two dimensions.  Further theoretical studies 
using this framework can be pursued exploring various properties of
multi-anyon systems including anyon interactions,
non-Abelian anyon correlations, and higher-order multi-anyon correlations.

\begin{acknowledgments}
The author would like to thank D.A. Brown, J.L. Klay, S.R. Klein, J. Randrup, R. Vogt, and Z. Yusof for valuable input
and stimulating discussions.
\end{acknowledgments}

\bibliography{anyon}

\begin{thebibliography}{30}
\expandafter\ifx\csname natexlab\endcsname\relax\def\natexlab#1{#1}\fi
\expandafter\ifx\csname bibnamefont\endcsname\relax
  \def\bibnamefont#1{#1}\fi
\expandafter\ifx\csname bibfnamefont\endcsname\relax
  \def\bibfnamefont#1{#1}\fi
\expandafter\ifx\csname citenamefont\endcsname\relax
  \def\citenamefont#1{#1}\fi
\expandafter\ifx\csname url\endcsname\relax
  \def\url#1{\texttt{#1}}\fi
\expandafter\ifx\csname urlprefix\endcsname\relax\def\urlprefix{URL }\fi
\providecommand{\bibinfo}[2]{#2}
\providecommand{\eprint}[2][]{\url{#2}}

\bibitem[{\citenamefont{Halperin}(1984)}]{fqhhal84k}
\bibinfo{author}{\bibfnamefont{B.}~\bibnamefont{Halperin}},
  \bibinfo{journal}{Phys.\ Rev.\ Lett.} \textbf{\bibinfo{volume}{52}},
  \bibinfo{pages}{1583} (\bibinfo{year}{1984}).

\bibitem[{\citenamefont{D.~Arovas and Wilczek}(1984)}]{fqharo84}
\bibinfo{author}{\bibfnamefont{J.~S.} \bibnamefont{D.~Arovas}}
  \bibnamefont{and} \bibinfo{author}{\bibfnamefont{F.}~\bibnamefont{Wilczek}},
  \bibinfo{journal}{Phys.\ Rev.\ Lett.} \textbf{\bibinfo{volume}{53}},
  \bibinfo{pages}{722} (\bibinfo{year}{1984}).

\bibitem[{\citenamefont{Laughlin}(1988)}]{alau88}
\bibinfo{author}{\bibfnamefont{R.}~\bibnamefont{Laughlin}},
  \bibinfo{journal}{Phys.\ Rev.\ Lett.} \textbf{\bibinfo{volume}{60}},
  \bibinfo{pages}{2677} (\bibinfo{year}{1988}).

\bibitem[{\citenamefont{Plyushchay}(1997)}]{suply97}
\bibinfo{author}{\bibfnamefont{M.~S.} \bibnamefont{Plyushchay}},
  \bibinfo{journal}{Mod. Phys. Lett.} \textbf{\bibinfo{volume}{A12}},
  \bibinfo{pages}{1153} (\bibinfo{year}{1997}).

\bibitem[{\citenamefont{Rausch~de Traubenberg and Slupinski}(1997)}]{surau97}
\bibinfo{author}{\bibfnamefont{M.}~\bibnamefont{Rausch~de Traubenberg}}
  \bibnamefont{and} \bibinfo{author}{\bibfnamefont{M.~J.}
  \bibnamefont{Slupinski}}, \bibinfo{journal}{Mod. Phys. Lett.}
  \textbf{\bibinfo{volume}{A12}}, \bibinfo{pages}{3051} (\bibinfo{year}{1997}).

\bibitem[{\citenamefont{Kitaev}(2003)}]{qckit03}
\bibinfo{author}{\bibfnamefont{A.~Y.} \bibnamefont{Kitaev}},
  \bibinfo{journal}{Ann. Phys.} \textbf{\bibinfo{volume}{303}},
  \bibinfo{pages}{2} (\bibinfo{year}{2003}).

\bibitem[{\citenamefont{Clark~{\em et al.}}(1988)}]{aclar88}
\bibinfo{author}{\bibfnamefont{R.}~\bibnamefont{Clark~{\em et al.}}},
  \bibinfo{journal}{Phys.\ Rev.\ Lett.} \textbf{\bibinfo{volume}{60}},
  \bibinfo{pages}{1747} (\bibinfo{year}{1988}).

\bibitem[{\citenamefont{Simmons~{\em et al.}}(1989)}]{asim89}
\bibinfo{author}{\bibfnamefont{J.}~\bibnamefont{Simmons~{\em et al.}}},
  \bibinfo{journal}{Phys.\ Rev.\ Lett.} \textbf{\bibinfo{volume}{63}},
  \bibinfo{pages}{1731} (\bibinfo{year}{1989}).

\bibitem[{\citenamefont{Wu}(1984)}]{homwu84}
\bibinfo{author}{\bibfnamefont{Y.-S.} \bibnamefont{Wu}},
  \bibinfo{journal}{Phys.\ Rev.\ Lett.} \textbf{\bibinfo{volume}{52}},
  \bibinfo{pages}{2103} (\bibinfo{year}{1984}).

\bibitem[{\citenamefont{Wilczek}(1982{\natexlab{a}})}]{awil4882}
\bibinfo{author}{\bibfnamefont{F.}~\bibnamefont{Wilczek}},
  \bibinfo{journal}{Phys.\ Rev.\ Lett.} \textbf{\bibinfo{volume}{48}},
  \bibinfo{pages}{1144} (\bibinfo{year}{1982}{\natexlab{a}}).

\bibitem[{\citenamefont{Wilczek}(1982{\natexlab{b}})}]{awil4982}
\bibinfo{author}{\bibfnamefont{F.}~\bibnamefont{Wilczek}},
  \bibinfo{journal}{Phys.\ Rev.\ Lett.} \textbf{\bibinfo{volume}{49}},
  \bibinfo{pages}{952} (\bibinfo{year}{1982}{\natexlab{b}}).

\bibitem[{\citenamefont{Aharonov and Bohm}(1959)}]{ab}
\bibinfo{author}{\bibfnamefont{Y.}~\bibnamefont{Aharonov}} \bibnamefont{and}
  \bibinfo{author}{\bibfnamefont{D.}~\bibnamefont{Bohm}},
  \bibinfo{journal}{Phys.\ Rev.} \textbf{\bibinfo{volume}{115}},
  \bibinfo{pages}{485} (\bibinfo{year}{1959}).

\bibitem[{\citenamefont{Leinaas and Myrheim}(1977)}]{alein77}
\bibinfo{author}{\bibfnamefont{J.}~\bibnamefont{Leinaas}} \bibnamefont{and}
  \bibinfo{author}{\bibfnamefont{J.}~\bibnamefont{Myrheim}},
  \bibinfo{journal}{Nuovo Cim.} \textbf{\bibinfo{volume}{B37}},
  \bibinfo{pages}{1} (\bibinfo{year}{1977}).

\bibitem[{\citenamefont{G.A.~Goldin and Sharp}(1980)}]{agold80}
\bibinfo{author}{\bibfnamefont{R.~M.} \bibnamefont{G.A.~Goldin}}
  \bibnamefont{and} \bibinfo{author}{\bibfnamefont{D.}~\bibnamefont{Sharp}},
  \bibinfo{journal}{J.\ Math.\ Phys.} \textbf{\bibinfo{volume}{21}},
  \bibinfo{pages}{650} (\bibinfo{year}{1980}).

\bibitem[{\citenamefont{G.A.~Goldin and Sharp}(1981)}]{agold81}
\bibinfo{author}{\bibfnamefont{R.~M.} \bibnamefont{G.A.~Goldin}}
  \bibnamefont{and} \bibinfo{author}{\bibfnamefont{D.}~\bibnamefont{Sharp}},
  \bibinfo{journal}{J.\ Math.\ Phys.} \textbf{\bibinfo{volume}{22}},
  \bibinfo{pages}{1664} (\bibinfo{year}{1981}).

\bibitem[{\citenamefont{Forte}(1992)}]{afor92}
\bibinfo{author}{\bibfnamefont{S.}~\bibnamefont{Forte}}, \bibinfo{journal}{Rev.
  Mod. Phys.} \textbf{\bibinfo{volume}{64}}, \bibinfo{pages}{193}
  (\bibinfo{year}{1992}).

\bibitem[{\citenamefont{Khare}(1997)}]{akhare1}
\bibinfo{author}{\bibfnamefont{A.}~\bibnamefont{Khare}},
  \emph{\bibinfo{title}{Fractional Statistics and Quantum Theory}}
  (\bibinfo{publisher}{World Scientific}, \bibinfo{year}{1997}).

\bibitem[{\citenamefont{Overbosch and Bais}(2001)}]{aover01}
\bibinfo{author}{\bibfnamefont{B.~J.} \bibnamefont{Overbosch}}
  \bibnamefont{and} \bibinfo{author}{\bibfnamefont{F.~A.} \bibnamefont{Bais}},
  \bibinfo{journal}{Phys. Rev.} \textbf{\bibinfo{volume}{A64}},
  \bibinfo{pages}{062107} (\bibinfo{year}{2001}).

\bibitem[{\citenamefont{Anchishkin
  et~al.}(2000{\natexlab{a}})\citenamefont{Anchishkin, Gavrilik, and
  Iorgov}}]{qbanch00a}
\bibinfo{author}{\bibfnamefont{D.~V.} \bibnamefont{Anchishkin}},
  \bibinfo{author}{\bibfnamefont{A.~M.} \bibnamefont{Gavrilik}},
  \bibnamefont{and} \bibinfo{author}{\bibfnamefont{N.~Z.}
  \bibnamefont{Iorgov}}, \bibinfo{journal}{Mod. Phys. Lett.}
  \textbf{\bibinfo{volume}{A15}}, \bibinfo{pages}{1637}
  (\bibinfo{year}{2000}{\natexlab{a}}).

\bibitem[{\citenamefont{Anchishkin
  et~al.}(2000{\natexlab{b}})\citenamefont{Anchishkin, Gavrilik, and
  Iorgov}}]{qbanch00b}
\bibinfo{author}{\bibfnamefont{D.~V.} \bibnamefont{Anchishkin}},
  \bibinfo{author}{\bibfnamefont{A.~M.} \bibnamefont{Gavrilik}},
  \bibnamefont{and} \bibinfo{author}{\bibfnamefont{N.~Z.}
  \bibnamefont{Iorgov}}, \bibinfo{journal}{Eur. Phys. J.}
  \textbf{\bibinfo{volume}{A7}}, \bibinfo{pages}{229}
  (\bibinfo{year}{2000}{\natexlab{b}}).

\bibitem[{\citenamefont{Haldane}(1991)}]{qhald91}
\bibinfo{author}{\bibfnamefont{F.~D.~M.} \bibnamefont{Haldane}},
  \bibinfo{journal}{Phys. Rev. Lett.} \textbf{\bibinfo{volume}{67}},
  \bibinfo{pages}{937} (\bibinfo{year}{1991}).

\bibitem[{\citenamefont{Hanbury~Brown and Twiss}(1956)}]{hbt}
\bibinfo{author}{\bibfnamefont{R.}~\bibnamefont{Hanbury~Brown}}
  \bibnamefont{and} \bibinfo{author}{\bibfnamefont{R.~Q.} \bibnamefont{Twiss}},
  \bibinfo{journal}{Nature} \textbf{\bibinfo{volume}{178}},
  \bibinfo{pages}{1046} (\bibinfo{year}{1956}).

\bibitem[{\citenamefont{Henny~{\em et al.}}(1999)}]{hbthen99}
\bibinfo{author}{\bibfnamefont{M.}~\bibnamefont{Henny~{\em et al.}}},
  \bibinfo{journal}{Science} \textbf{\bibinfo{volume}{284}},
  \bibinfo{pages}{296} (\bibinfo{year}{1999}).

\bibitem[{\citenamefont{Goldhaber et~al.}(1960)\citenamefont{Goldhaber,
  Goldhaber, Lee, and Pais}}]{hbtgold60}
\bibinfo{author}{\bibfnamefont{G.}~\bibnamefont{Goldhaber}},
  \bibinfo{author}{\bibfnamefont{S.}~\bibnamefont{Goldhaber}},
  \bibinfo{author}{\bibfnamefont{W.-Y.} \bibnamefont{Lee}}, \bibnamefont{and}
  \bibinfo{author}{\bibfnamefont{A.}~\bibnamefont{Pais}},
  \bibinfo{journal}{Phys. Rev.} \textbf{\bibinfo{volume}{120}},
  \bibinfo{pages}{300} (\bibinfo{year}{1960}).

\bibitem[{\citenamefont{Boal et~al.}(1990)\citenamefont{Boal, Gelbke, and
  Jennings}}]{hbtboal90}
\bibinfo{author}{\bibfnamefont{D.~H.} \bibnamefont{Boal}},
  \bibinfo{author}{\bibfnamefont{C.~K.} \bibnamefont{Gelbke}},
  \bibnamefont{and} \bibinfo{author}{\bibfnamefont{B.~K.}
  \bibnamefont{Jennings}}, \bibinfo{journal}{Rev. Mod. Phys.}
  \textbf{\bibinfo{volume}{62}}, \bibinfo{pages}{553} (\bibinfo{year}{1990}).

\bibitem[{\citenamefont{Baym}(1998)}]{hbtbay98}
\bibinfo{author}{\bibfnamefont{G.}~\bibnamefont{Baym}}, \bibinfo{journal}{Acta
  Phys. Polon.} \textbf{\bibinfo{volume}{B29}}, \bibinfo{pages}{1839}
  (\bibinfo{year}{1998}), \eprint{nucl-th/9804026}.

\bibitem[{\citenamefont{Scully and Zubairy}(1997)}]{qoscull1}
\bibinfo{author}{\bibfnamefont{M.~O.} \bibnamefont{Scully}} \bibnamefont{and}
  \bibinfo{author}{\bibfnamefont{M.~S.} \bibnamefont{Zubairy}},
  \emph{\bibinfo{title}{Quantum Optics}} (\bibinfo{publisher}{Cambridge
  University Press}, \bibinfo{year}{1997}).

\bibitem[{\citenamefont{Koonin}(1977)}]{hbtkoo77}
\bibinfo{author}{\bibfnamefont{S.~E.} \bibnamefont{Koonin}},
  \bibinfo{journal}{Phys. Lett.} \textbf{\bibinfo{volume}{B70}},
  \bibinfo{pages}{43} (\bibinfo{year}{1977}).

\bibitem[{\citenamefont{Pratt et~al.}(1990)\citenamefont{Pratt, Csoergoe, and
  Zimanyi}}]{hbtpra90}
\bibinfo{author}{\bibfnamefont{S.}~\bibnamefont{Pratt}},
  \bibinfo{author}{\bibfnamefont{T.}~\bibnamefont{Csoergoe}}, \bibnamefont{and}
  \bibinfo{author}{\bibfnamefont{J.}~\bibnamefont{Zimanyi}},
  \bibinfo{journal}{Phys. Rev.} \textbf{\bibinfo{volume}{C42}},
  \bibinfo{pages}{2646} (\bibinfo{year}{1990}).

\bibitem[{\citenamefont{Brown and Danielewicz}(1997)}]{hbtbro97}
\bibinfo{author}{\bibfnamefont{D.~A.} \bibnamefont{Brown}} \bibnamefont{and}
  \bibinfo{author}{\bibfnamefont{P.}~\bibnamefont{Danielewicz}},
  \bibinfo{journal}{Phys. Lett.} \textbf{\bibinfo{volume}{B398}},
  \bibinfo{pages}{252} (\bibinfo{year}{1997}).

\end{thebibliography}

\end{document}